\documentclass[aps,prd,
superscriptaddress,groupedaddress,nofootinbib]{revtex4} 
\usepackage{amsmath,amssymb}
\usepackage{epsfig,graphicx}
\usepackage{amssymb}
\usepackage{epsfig,color}
\usepackage{pstricks,graphicx,epsfig,color,amssymb,amsmath,amscd}
\usepackage{subfigure}
\graphicspath{
	{./Pics/}
}

\usepackage{braket}
\usepackage{newunicodechar}
\newunicodechar{⊙}{}

\usepackage{makeidx}
\newcommand{\bi}{\bibitem}
\newcommand{\be}{\begin{eqnarray}}
\newcommand{\ee}{\end{eqnarray}}
\newcommand{\rar}{\rightarrow}

\usepackage[]{caption}
\def\-g{\sqrt{-g}}
\renewcommand\rho{\varrho}

\usepackage{multirow}

\begin{document}

\title{\bf Baryogenesis through baryon capture by black holes 
}
\author{A.~D.~Dolgov}
\email{dolgov@fe.infn.it}
\affiliation{	Novosibirsk State University \\
	Pirogova ul., 2, 630090 Novosibirsk, Russia  }

\author{N.~A.~Pozdnyakov }
\email{pozdniko@gmail.com}
\affiliation{	Novosibirsk State University \\
	Pirogova ul., 2, 630090 Novosibirsk, Russia  }

\date{\today}

\begin{abstract}
A novel mechanism of cosmological baryogenesis through baryon capture by primordial black holes is suggested. In 
contrast to the conventional scenarios  it does not demand  non-conservation of baryonic number in particle physics 
and can proceed in thermal equilibrium. For implementation of  this mechanism a heavy superweakly interacting  
particle a with non-zero baryon number is necessary.
\end{abstract}


\maketitle


\section{Introduction \label{s-intro}}
 
The explanation of the observed excess of baryons over antibaryons in the universe was beautifully solved by 
Sakharov~\cite{ADS-BG} in 1967.  He formulated the following three necessary conditions for generation of the 
cosmological baryon asymmetry:

1. Violation of C and CP symmetries in particle physics.

2. Non-conservation of baryonic number, $B$.

3. Deviation from thermal equilibrium in the early universe.\\
With properly chosen parameters of the particle physics model at high energies these three principles allow to
explain the cosmological excess of particles over antiparticles and to calculate the magnitude of the asymmetry.
According to the review~\cite{pdg}, the magnitude of the asymmetry, expressed in terms of the present day
number densities,  is equal to 
\be 
 \beta = \frac{ n_B - n_{ \bar B}} { n_\gamma}  \approx 6 \cdot 10^{-10}   
\label{beta}
\ee
where  $n_B$ and  $n_{\bar B}$  are respectively the number densities of baryons and antibaryons 
(note that today $n_{\bar B} \ll n_B)$,
$n_\gamma = 411(T_\gamma /2.73^o {\rm K})^3 {\rm  cm^{-3}}$, and $T_\gamma = 2.73^o {\rm K}$ is the 
present day temperature of the cosmic microwave background (CMB) radiation.
 
There exists a  plethora of different models of baryogenesis, which can successfully do the job,
for reviews see refs.~\cite{AD-YaZ,ad1,ckn1,rs,ad2,RiTr:RPiB,jmc}. Here we consider rather special scenario 
invoking black holes for creation of cosmological baryon asymmetry. The idea that black hole evaporation
can lead to different numbers of particles and antiparticles in outer world belongs to Hawking~\cite{sh-bg}
and Zeldovich~\cite{yaz-bg} and was quantitatively realized in refs.~\cite{ad-bh1,ad-bh2}. 

Briefly the mechanism operates as follows. The evaporated particles initially have thermal equilibrium distribution
and there should be equal numbers of particles and antiparticles. However, they propagate in the gravitational
field of the parent BH with the effective potential barrier at a distance of several gravitational radii. The height 
of the barrier is different for particles of different masses, higher for larger mass. {Hence} the presence of the 
barrier breaks the equilibrium. Assume that among the evaporated particles there are some massive 
bosons which can decay 
into a pair of quark and antiquark of different  flavor, e.g. into $u$ and $\bar t$ or vice versa into the 
charge conjugated channel, i.e. into $\bar u$ and $t$.
Assume also that due to breaking of C and CP symmetries the probability of the decay into $u \bar t$ is
larger than into $ \bar t u$, then the flux of  baryons into external space would be larger than the flux of antibaryons.

Here we consider in a sense opposite scenario of different absorption rate of  baryon and antibaryons by a BH. 
This mechanism may generate cosmological baryon asymmetry in thermal equilibrium without breaking of
B-conservation. In the following section we discuss the capture mechanism of $X/\bar X$ particles by PBH. 
In the third section the difference  between $X$ and $\bar X$ scattering cross sections is estimated.
In sec. IV we conclude.

\section{Capture mechanism by PBH {in the early universe} \label{s-mob}}

Assume that there exist heavy particles $X$ and antiparticles $\bar X$ with non-zero baryon number $B$. 
They are supposed to be unstable, but possibly long-lived,
decaying into {light particles through}  
$B$-conserving decays. Let us consider the epoch of the cosmological evolution when the temperature was of the order of 
the $X$-particle mass, $T \approx m_X$, while all other particles have zero or much smaller masses. Also assume that 
there exists an interaction between $X$, $\bar X$ particles and light particles (e.g. quarks) which respects
CPT symmetry but breaks CP and C symmetries.

Assume also that in the early universe there existed a population of primordial black holes
with the energy density at  the moment of their creation much smaller than the energy density of the 
background particles, 
\be
\rho_{BH} = \epsilon \rho_{0},
\label{rho_BH}
\ee
with $\epsilon \ll 1$.

According to the suggested scenario, heavy particles $X$ and $\bar X$ have been captured by PBHs with
some excess of $\bar X$ over $X$ due to a small difference of the interaction strength of $X$ and $\bar X$
with cosmological plasma.

The equations of motion for  $X$ and $\bar X$ near the PBH can be written in analogy with those of ref.~\cite{CB-AD-AP}
and have the form:
\be
\dot v_X &=& -\frac{G_N M}{r^2} 
+ \frac{\alpha Q}{r^2 m_X} 
+ \frac{L \, \sigma_{0 X}}{4 \pi r^2 m_X} 
- \frac{\sigma_{0 X} n_0 E_0}{m_X} \, v_X
- \frac{n_X \sigma_{X \bar X} \delta P}{m_X} \, (v_X - v_{\bar X}) \, 
\label{dot-v-X} \\
\dot v_{\bar X} &=& -\frac{G_N M}{r^2} 
- \frac{\alpha Q}{r^2 m_X} 
+ \frac{L \, \sigma_{0 \bar X}}{4 \pi r^2 m_X} 
- \frac{\sigma_{0 \bar X} n_0 E_0}{m_X} \, v_{\bar X}
+ \frac{n_{\bar X} \sigma_{X \bar X} \delta P}{m_X} \, (v_X - v_{\bar X}) .
\label{dot-v-barX} 
\ee
Here $M$ is the PBH mass, $G_N = 1/m_{Pl}^2$ is gravitational constant, $m_{Pl} = 1.22 \cdot 10^{19}$ GeV is the Planck mass,
$v_X$ and $v_{\bar X}$ are the fluid velocities of $X$ and $\bar X$
(that is the average velocities of $X$ and $\bar X$ in the plasma 
which do not include typically larger chaotic thermal velocities), $Q$ is the electric charge of the PBH in 
elementary charge units,  $\alpha = e^2/4\pi = 1/137$, $\sigma_{ij}$ is the cross section of 
scattering of $i$ on $j$,  sub-index $0$ means all the set of light relativistic particles in the 
primeval plasma, with masses much smaller than $m_X$;
$L$ is the luminosity induced by the accretion to the PBH flow, $\delta P$ is the momentum transfer in 
$X \bar X$--scattering, $n_X$ and $n_{\bar X}$ are the number densities of $X$ and $\bar X$ 
around the PBH, $n_0$ is the number density of the light particles in thermal bath surrounding the PBH, 
and  $E_0$ is the light particle energy, which is roughly equal to the momentum transfer 
in $X $ and $\bar X$ scattering on light particles. We neglect the angular
momentum term in the equations above
because it does not change appreciably  our results.

We are interested in the difference of the capture velocities of $X$ and $\bar X$. The overall motion of 
the light particles towards PBHs are disregarded. The account of this motion would enhance the capture 
probability of heavy ones, $X$ and $\bar X$.

It is convenient to introduce the following quantities: 
\be
v_X + v_{\bar X} &=& 2 v_+, \,\,\, v_X - v_{\bar X} =  v_- \nonumber\\
\sigma_{0X} + \sigma_{0\bar X} &=& 2 \sigma_+, \,\,\, \sigma_{0X} - \sigma_{0\bar X} =  \sigma_-  ,
\label{v-plus}
\ee
In the first order in $v_-$ and $\sigma_- $  equations (\ref{dot-v-X}) and (\ref{dot-v-barX}) turn into:
\be
\dot v_+ &=& -\frac{G_N M}{r^2} 
+ \frac{L \, \sigma_{+}}{4 \pi r^2 m_X} 
- \frac{\sigma_{+} n_0 E_0}{m_X} \, v_+ ,
\label{dot-v-plus} \\
\dot v_{-} &=& 
\frac{2\alpha Q}{r^2 m_X} 
+ \frac{L \, \sigma_{-}}{4 \pi r^2 m_X}  
- \frac{ n_0 E_0}{m_X} \left( \sigma_+ v_{-} +  \sigma_- v_{+} \right)
- (n_X + n_{\bar X}) \frac{\sigma_{X \bar X} \delta P}{m_X} \,v_-  .
\label{dot-v-minus} 
\ee
The acquired by PBH electric charge $Q$ would be quickly neutralized by light charged particles so it is
neglected in what follows. 

We assume that the characteristic time is smaller than the Hubble time, so the coefficients in the above 
equations can be treated as constant and thus equation (\ref{dot-v-plus}) is solved as
\be
v_+ = \left(  \frac{L \sigma_+}{4\pi m_X} -  \frac{M}{m_{Pl}^2} \right)
\cdot \frac{m_X \left(1 - e^{-\gamma_+ t }\right)}{r^2 \sigma_+ n_0 E_0} 
\rar \left( \frac{L \sigma_+}{4\pi m_X}  - \frac{M}{m_{Pl}^2}  \right)
\cdot \frac{m_X }{r^2 \sigma_+ n_0 E_0} ,
\label{v-plus}
\ee
where $\gamma_+ = \sigma_+ n_0 E_0/m_X$. For large $t \gamma_+$ we can neglect $e^{-\gamma_+ t} $
in comparison with 1.

With known $v_+$ eq.~(\ref{dot-v-minus}) can be solved for large $t$ as
\be
v_- = \frac{M}{m_{Pl}^2 r^2}\,\frac{\sigma_-}{\sigma_+}  \frac{m_X}{n_0 E_0 \sigma_+}
 \left( 1 + \frac{n_X + n_{\bar X}}{n_0} \frac{\sigma_{X \bar X}}{\sigma_+}\frac{\delta E_0}{E_0} \right)^{-1}
 \approx \frac{M}{m_{Pl}^2 r^2}\,\frac{\sigma_-}{\sigma_+} \frac{m_X}{n_0 E_0 \sigma_+}.
\label{v-minus}
\ee

The (anti)baryonic number could be accumulated inside PBH during the Hubble time when the cosmic temperature
was close to $m_X$. At that period there was: $n_0 = 0.1 g_* T^3 \approx 10 m_X^3$, 
$n_X = k n_0$ with $k= 0.01-0.1$,
$E_0 \approx m_X$ and  $\sigma_+ = f^4 g_*/m_X^2$, 
where $g_* \approx 100$ is the number of particle species in 
the cosmic plasma, $f$ is a small coupling constant of $X$-particles to other ones. The Hubble time at that period
was:
\be
t_H = \frac{1}{2H}  = \left(\frac{90}{32 \pi^3 g_*}\right)^{1/2} \frac{m_{Pl} }{T^2} \approx 0.03 \, \frac{m_{Pl} }{m_X^2}.
\label{t-H}
\ee   
The total baryonic number of PBH collected from the volume inside the sphere of radius $r$ during $t=t_H$ 
would be:
\be  
 N_B = 4 \pi r^2 t_H n_X v_- B_X \approx \frac{k}{0.01} f^{-2} B_X  M_g,
 \label{N-B}
 \ee
 where $M_g$ if the PBH mass in grams, $B_X$ is a baryon number of $X$ particle, and $\sigma_-/\sigma_+ \sim f^2$, because the difference between
 cross-sections of particles and antiparticles due to breaking of C and CP symmetries arises in the lowest order
 loop corrections in the same way as the difference between partial decay width of particles and antiparticles
 appears due to rescattering in the final state as it is discussed in reviews on baryogenesis quoted above, 
 especially in~\cite{AD-YaZ}. 
 
 The baryonic asymmetry $\beta = n_B/n_\gamma$, where $n_B$ is the density of the baryonic number and
 $n_\gamma$ is the present day number density of cosmic microwave background radiation, can be estimated 
 as
 \be
 \beta = \frac{n_{BH}}{n_0}  N_B = \frac{\rho_{BH}}{M} \frac{3 T}{\rho_0} N_B 
 \approx 5 \cdot 10^{-24} B_X  \frac{\epsilon}{ f^{2}} \,\frac{m_X }{{\rm GeV}},
 \label{beta}
 \ee
assuming that this ratio remains constant in the course of cosmological evolution. This is approximately true
if the entropy rise due to annihilation of massive particle species is not too strong. In the standard model the
entropy dilution is equal to the ratio of particle species from $T= m_X$ down to a fraction of MeV. It dilutes 
$\beta$ roughly by factor 10. However, an abundant population of evaporating PBHs could lead to considerable dilution. 
According to the calculations of ref.~\cite{ac-ad} the dilution factor is
\be
S = 10^5 \epsilon M_g ,
\label{S}
\ee
 which we demand to be not much larger than unity. 
 
 If the coupling constant $f^2$ is of the order of the fine structure constant $\alpha \sim 0.01$, it is difficult
 to generate the observable asymmetry $\beta \sim 10^{-9}$ through $X/\bar X$-capture by PBH,
 but  super-weakly interacting 
$f \ll \alpha$, and sufficiently massive $X$-particles could successfully do the job. 

{\section{Difference between mobilities of $X$ and $\bar X$ particles in the background plasma \label{s-mob}} }

As we have already mentioned,  C and CP symmetries are assumed to be broken, while CPT remains intact.
As it is well known out of these three symmetries only CPT has rigorous theoretical justification. 
Namely its validity follows from the locality of particle interactions, Lorentz invariance, and hermicity of the 
Hamiltonian. 
 
CPT invariance implies equality of the particle and antiparticle masses and their total decay widths, while  the partial decay
widths of particles and antiparticles into charge conjugate channels may be different, due to breaking of  C and CP. 
However, it can happen only in higher orders of perturbatiion theory.
Similar inequality may also exist for probabilities of particle and antiparticle scattering:
\be
\sum_{a,b} \Gamma ( X +a \rar X + b) \neq \sum_{\bar a, \bar b} \Gamma ( \bar X +\bar a \rar \bar X + \bar b),
\label{sigma-q-barq}
\ee
where summations are done over all  light particle sets in initial, $a$, and final states, $b$. 

However, we should take into account that according to CPT theorem the total probability of a 
process from an initial state, containing  a certain set of particles 
is equal to the total probability of the process containing antiparticles with opposite spin projection state, e.g.
with opposite helicities $\lambda$~\cite{AD-YaZ,SW-QFT}:

\be
\Gamma \left[ {\bf p}_1, \lambda_1, a_1, {\bf p}_2, \lambda_2, a_2, ... \right] =
\Gamma \left[ {\bf p}_1, -\lambda_1, \bar a_1, {\bf p}_2, -\lambda_2, \bar a_2, ... \right] 
\label{Gamma-tot}
\ee

In particular, this condition leads to the mentioned above equality of the total decay width of particles and antiparticles while allows for
a difference of among the partial decay rates.

To achieve desired difference of $X$ and $\bar X$ particles scattering 
 we need to introduce a new interaction leading to
disappearance of $X$ and $\bar X$ particles through decay or scattering process, such as
\be
a + X \rar b + H,
\label{AX}
\ee
where the heavy particle $H$ has zero baryonic number and
the light state $b$ should have the same baryonic number as 
	$a + X$ if we want to avoid
non-conservation of baryons.

\begin{figure}[ht]
	 \centering \subfigure[]{
		\includegraphics[width=0.25\linewidth]{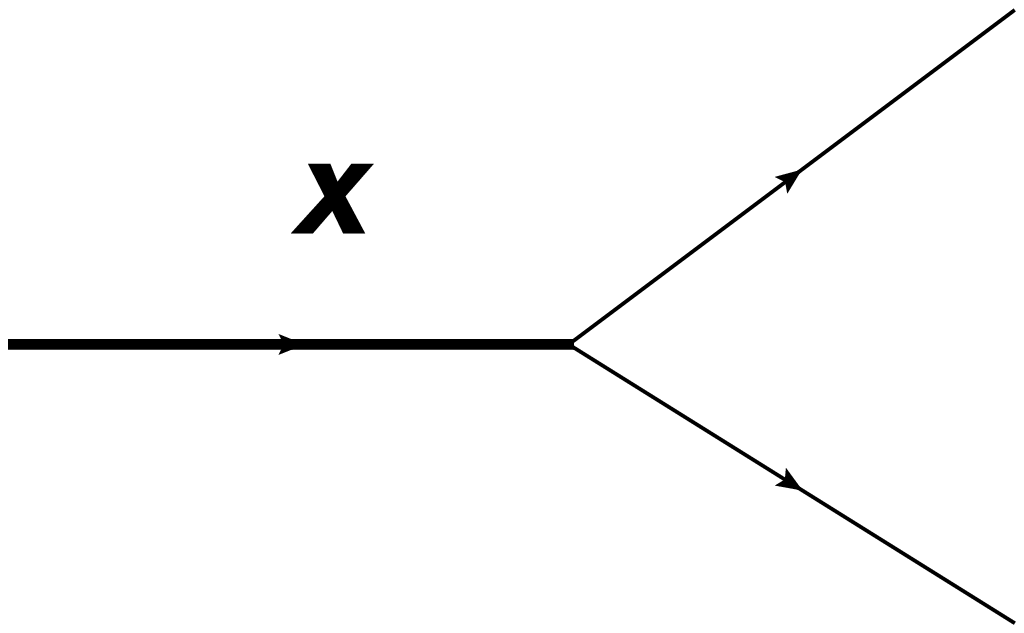}
		\label{fig:Xdecay} }
	\hspace{4ex}
	\subfigure[]{
		\includegraphics[width=0.25\linewidth]{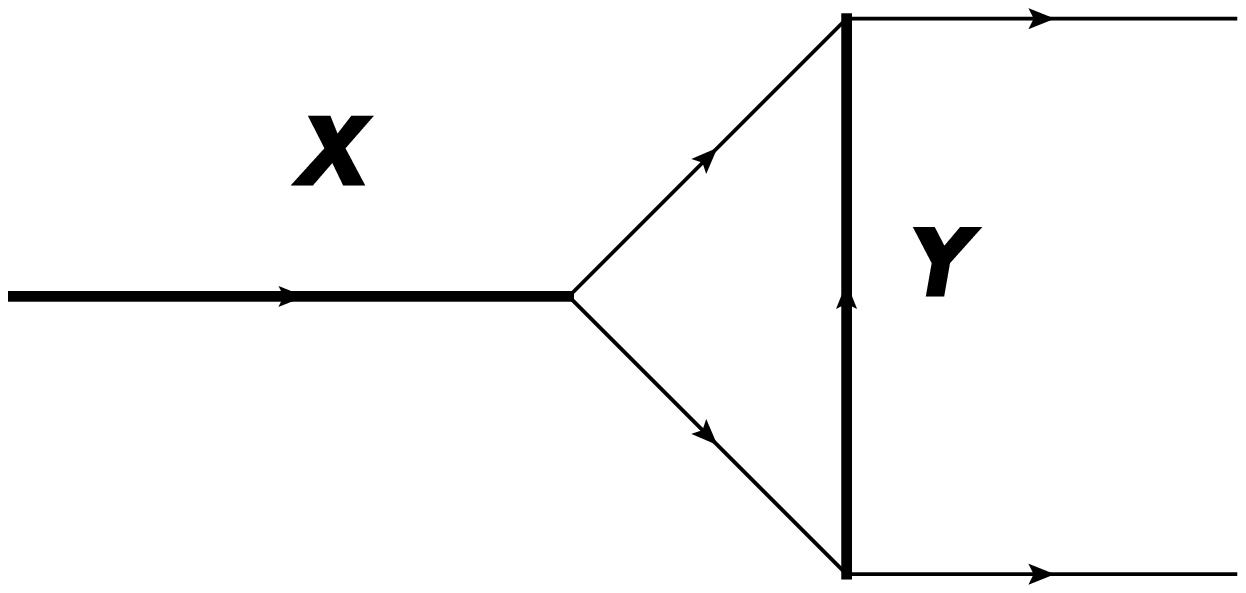}
		\label{fig:interf} }
	\caption{{Feynman diagrams describing the X-particle decay in the lowest order
		(a) and loop correction with rescattering process (b).}
	}
	\label{fig:1}	
\end{figure}

{It is instructive to consider an example of how a difference between partial decay rates appears} in the case 
of two-body decays of X-particle. A detailed study of the
necessary number of the decay channels and the number of different X-like particle species is   
presented in ref.~\cite{AD-YaZ}. In the lowest order in perturbation theory the decay is described by
the {diagram presented in figure \ref{fig:Xdecay}}. Evidently in this order the probability of charge conjugated decays   
are equal due to hermicity of the Lagrangian. So the amplitudes of particle and antiparticle decays have 
opposite phases but equal absolute values. 
To avoid this limitation we need add the higher order correction
depicted in figure \ref{fig:interf}.
This higher order correction to the amplitude has an additional imaginary part due to on-mass-shell
re-scattering in the final state through an exchange of another heavy particle $Y$, usually different from $X$.

Typically the decay width with one-loop correction taken into account 
is given  by the expression of the kind:
\be 
\Gamma (X \rar a+b) = \alpha c_0 m_X (1 + \alpha c_1 ),
\label{X-to-ab}
\ee
where $c_0$ and $c_1$ are numerical constants generically of the order of unity; $c_0$ is the same for particles and
antiparticles, while $c_1$ is different if C and CP are broken. However, as we have already mentioned, the sum over
all final states produced in the decay are the same for $X$ and $\bar X$.  
The difference in the partial decay widths is a cornerstone of the popular model of 
baryogenesis by heavy particle decays.


\begin{figure}[ht]
	\centering \subfigure[]{
		\includegraphics[width=0.25\linewidth]{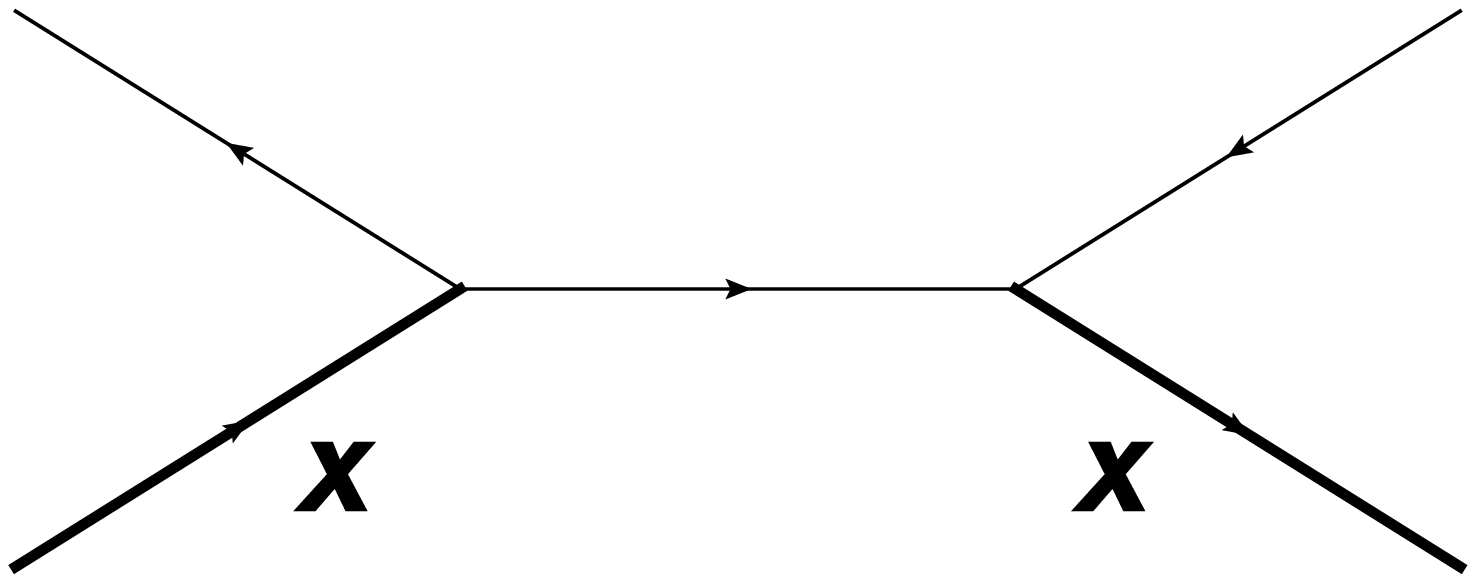}
		\label{fig:Xsc} }
	\hspace{4ex}
	\subfigure[]{
		\includegraphics[width=0.25\linewidth]{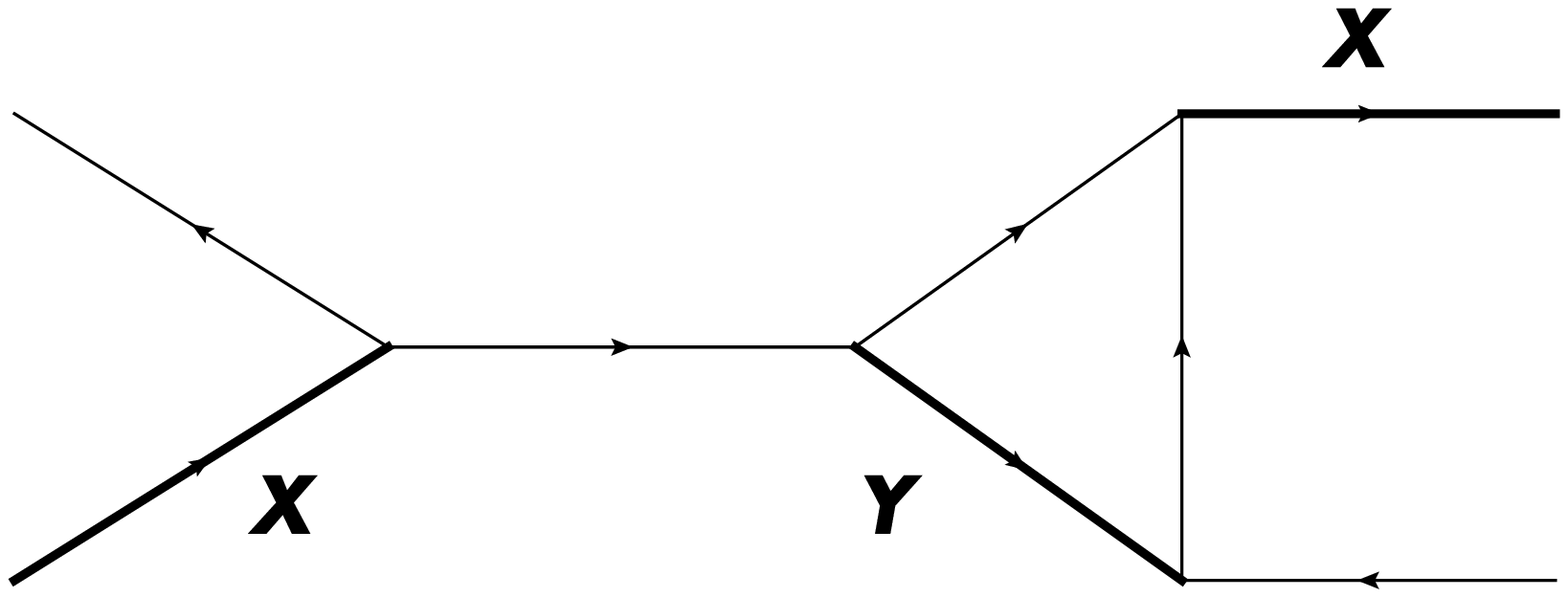}
		\label{fig:Xresc} }
	\caption{{Feynman diagrams describing $X$ (or $\bar X$)
			scattering off light quark. The first, 'a', diagram is
			the lowest order contribution and gives equal values for particles and antiparticles.  
			The 'b' diagram present an example of the one-loop correction with an exchange of $Y$-particle. 
			The one-loop scattering may be also present to the initial state and its contribuition multiplies the result
			by factor 2.
		}}
	\label{fig:2}	
\end{figure}

The equalities of the total probabilities means in particular that the total cross-section of $X$-scattering on a particle $"a"$
$\sigma_{tot} (X+a \rar All) $ is equal to the same of particles and antiparticles $\sigma_{tot} (\bar X+ \bar a \rar All) $. 
If the final state "All" contains one and only one $X$-particle, then the mobilities of $X$ and $\bar X$ in the cosmological
plasma would be the same and the discussed here mechanism of baryogenesis would not operate.
However, if the complete set of the final states include a state or states where $X$ (or $\bar X$) is missing, then 
the cross-sections of the processes $\sigma_{tot} (X+a \rar X+ All) $ and  $\sigma_{tot} (\bar X+ \bar a \rar  \bar X +All) $.
may be different, leading to the needed mobility differences of $X$ and $\bar X$.
In complete analogy with the differences of the widths of the charge conjugated decay channels
difference between the cross-sections appears due to radiative corrections to the lowest order amplitude, 
see fig.~\ref{fig:Xresc}. The difference between probabilities of charge conjugated processes can be estimated as:

\be
\delta = \frac{\sigma_{X0}-\sigma_{\bar X 0}}{\sigma_{X0}+\sigma_{\bar X 0}} \approx \frac{|g_{x1}|^2Im(D)Im(g_{x1}g_{x2}^*g_{y1}g_{y2}^*)}{|g_{x1}|^4} \propto \alpha ,
\ee
where the coefficient $D$ comes from the integration over the loop, 
$g_{xi}$ and $g_{yi}$ are partial decay constants of $X$ and $Y$ particles respectively.

The following supersymmetry  inspired model can serve as appropriate frameworks for the scenario.
Assume that $X$ is an analogue of the lightest supersymmetric particles (LSP)
which is stable due to an analogue of $R$-parity. Let assume that there exists
a heavier partner $H$ with zero baryonic number which would be unstable and decay through the 
channel $H \rar X+ 3 q$, where $q$ are light quarks with proper quantum numbers.
Accordingly the reaction $X + q \rar H + 2 \bar q$ becomes possible. It is exactly what we need to allow
for a difference between the cross-sections of the reaction $\sigma_{tot} (X+a \rar X+ All) $ and 
$\sigma_{tot} (\bar X+ \bar a \rar  \bar X +All) $, which can lead to a different mobilities of $X$ and $\bar X$
around black hole and to dominant capture of antibaryons over baryons creating cosmological baryon asymmetry.

Note that in $R^2$ gravity, Srarobinsky inflation~\cite{aas-infl}, the allowed mass of LSP-kind particle can
be up to $10^{13}$ GeV~\cite{EA-AD-RS-1,EA-AD-RS-2,EA-AD-RS-3}.

\section{Conclusion \label{s-concl} }

The proposed here scenario of baryogenesis has a novel feature that non-conservation of baryonic number  
in particle interactions is unnecessary. So the proton must be almost absolutely stable. To be more precise
it may decay by Zeldovich mechanism~\cite{zeld-BH}
 through formation of a virtual black hole from three quarks inside proton. But the life-time
 with respect to such decay is almost infinite, $\tau_p \sim 10^{45}$ years. Also one could hardly expect 
neutron-antineutron oscillations  to be observable (for a recent review see e.g. \cite{n-anti-n}).

Another unusual feature of the model is a possibility to create baryon (or any other type of asymmetry between
particles and antiparticles) in thermal equilibrium. Normally the suppression factor is of the order of the ratio
of the Hubble expansion rate to the particle expansion rate, $H/\Gamma$. The former is inversely proportional
to a huge value of the Planck mass, $H \sim T^2/m_{Pl}$, where $m_{Pl} = 1.22\cdot 10^{19}$ GeV and $T$ is
the cosmological plasma temperature. According to the estimates (\ref{t-H}) and (\ref{N-B}),
for the mechanism considered here the situation is opposite: the larger is the Planck mass (or the slower is the 
cosmological expansion), the larger is the baryon asymmetry.

\section*{Acknowledgment}

This work was supported by RNF Grant  19-42-02004. 

The Feynman diagrams was drawn by JaxoDraw~\cite{jaxo}.

\end{document}